\documentstyle[aps,epsfig,multicol]{revtex}
\begin{document}

\twocolumn
\twocolumn[
\hsize\textwidth\columnwidth\hsize\csname@twocolumnfalse\endcsname

\title{Importance of Correlation Effects on Magnetic Anisotropy in Fe and Ni }
\author{Imseok Yang, Sergej Y. Savrasov, and Gabriel Kotliar}
\address{Department of Physics and Astronomy and Center for Condensed Matter Theory,\\
Rutgers University, Piscataway, NJ 08854\\
}
\date{\today}
\maketitle

\begin{abstract}
We calculate magnetic anisotropy energy of Fe and Ni by taking into account
the effects of strong electronic correlations, spin-orbit coupling, and
non-collinearity of intra--atomic magnetization. The LDA+U method is used
and its equivalence to dynamical mean--field theory in the static limit is
emphasized. Both experimental magnitude of MAE and direction of
magnetization are predicted correctly near $U=1.9\ eV$, $J=1.2\ eV$ for Ni
and $U=1.2\ eV$, $J=0.8\ eV$ for Fe. Correlations modify the one--electron
spectra which are now in better agreement with experiments.
\end{abstract}


]

The calculation of the magneto-crystalline anisotropy energy (MAE)~\cite
{vanVleck:37,brooks,fletcher,sloncewskij,asdente} of magnetic materials
containing transition-metal elements from first principles electronic
structure calculations is a long-standing problem. The MAE is defined as the
difference of total energies with the orientations of magnetization pointing
in different, e.g., (001) and (111), crystalline axis. The difference is not
zero because of spin-orbit effect, which couples the magnetization to the
lattice, and determines the direction of magnetization, called the easy axis.

Being a ground state property, the MAE should be accessible in principle via
density functional theory (DFT)~\cite{hk64,ks65}. Despite the primary
difficulty related to the smallness of MAE ($\sim 1\ \mu eV/$atom), great
efforts to compute the quantity with advanced total energy methods based on
local density approximation (LDA) combined with the development of faster
computers, have seen success in predicting its correct orders of magnitudes~ 
\cite{efn,halilov,jansen,tjew,dks}. However, the correct easy axis of Ni has
not been predicted by this method and the fundamental problem of
understanding MAE is still open.

A great amount of work has been done to understand what is the difficulty in
predicting the correct axis for Ni. Parameters within the LDA calculation
have been varied to capture physical effects which might not be correctly
described. These include (i) scaling spin-orbit coupling in order to enlarge
its effect on the MAE~\cite{halilov,jansen}, (ii) calculating torque to
avoid comparing large numbers of energy~\cite{jansen}, (iii) studying the
effects of the second Hunds rule in the orbital polarization theory ~\cite
{tjew}, (iv) analyzing possible changes in the position of the Fermi level
by changing the number of valence electrons ~\cite{dks}, (v) using the state
tracking method~\cite{freeman}, and (iv) real space approach~\cite{beiden}.

In this paper we take a new view that the correlation effects within the $d$
shell are important for the magnetic anisotropy of $3d$ transition metals
like Ni. These effects are not captured by the LDA but are described by
Hubbard--like interactions presented in these systems and need to be treated
by first principles methods\cite{anisimov}.

Another effect which has not been investigated in the context of magnetic
anisotropy calculations is the non-collinear nature of intra-atomic
magnetization~\cite{singh}. It is expected to be important when spin-orbit
coupling and correlation effects come into play together. In this article we
show that when we include these new ingredients into the calculation we
solve the long-standing problem of predicting the correct easy axis of Ni.

We believe that the physics of transition metal compounds is intermediate
between atomic limit where the localized $d$ electrons are treated in the
real space and fully itinerant limit where the electrons are described by
band theory in k space. A many--body method incorporating these two
important limits is the dynamical mean--field theory (DMFT)~\cite{gabi:rmp}.
The DMFT approach has been extensively used to study model Hamiltonian of
correlated electron systems in the weak, strong and intermediate coupling
regimes. It has been very successful in describing the physics of realistic
systems such as transition metal oxides and, therefore, is expected to treat
properly the materials with $d$ or $f$ electrons.

Electron--electron correlation matrix $U_{\gamma _{1}\gamma _{2}\gamma
_{3}\gamma _{4}}=\left\langle m_{1}m_{3}\left| v_{C}\right|
m_{2}m_{4}\right\rangle \delta _{s_{1}s_{2}}\delta _{s_{3}s_{4}}$ for $d$
orbitals is the quantity which takes strong correlations into account. This
matrix can be expressed via Slater integrals $F^{(i)} $, $i=0,2,4,6$ in the
standard manner. The inclusion of this interaction generates self--energy $%
\Sigma_{\gamma_1\gamma_2}(i\omega_n,\vec{k})$ on top of the one--electron
spectra. Within DMFT it is approximated by momentum independent self--energy 
$\Sigma_{\gamma_1\gamma_2}(i\omega_n)$.

A central quantity of the dynamical mean--field theory is the one--electron
on--site Green function 
\begin{eqnarray}
G_{\gamma_1\gamma_2}(i\omega_n)=&\sum_{\vec{k}}&\left[(i\omega_n+\mu)
O_{\gamma_1\gamma_2}(\vec{k}) -H^0_{\gamma_1\gamma_2}(\vec{k})\right. 
\nonumber \\
&+&\left. v_{dc}-\Sigma_{\gamma_1\gamma_2} (i\omega_n)\right]^{-1}.
\label{dmft}
\end{eqnarray}
where $H^0_{\gamma_1\gamma_2}(\vec{k})$ is the one--electron Hamiltonian
standardly treatable within the LDA. Since the latter already includes the
electron-electron interactions in some averaged way, we subtract the double
counting term $v_{dc}$~\cite{laz}. The use of realistic localized orbital
representation such as linear muffin--tin orbitals \cite{OA75} leads us to
include overlap matrix $O_{\gamma_1\gamma_2}(\vec{k})$ into the calculation.

The DMFT reduces the problem to solving effective impurity model where the
correlated $d$ orbitals are treated as an impurity level hybridized with the
bath of conduction electrons. The role of hybridization is played by the
so--called bath Green function defined as follows: 
\begin{equation}
[{\cal G}_0^{-1}]_{\gamma_1\gamma_2}(i\omega_n)=
G_{\gamma_1\gamma_2}{}^{-1}(i\omega_n) +\Sigma_{\gamma_1\gamma_2}(i\omega_n).
\end{equation}
Solving this impurity model gives access to the self--energy $%
\Sigma_{\gamma_1\gamma_2}(i\omega_n)$ for the correlated electrons. The
one--electron Green function~(\ref{dmft}) is now modified with new $%
\Sigma_{\gamma_1\gamma_2}(i\omega_n)$, which generates a new bath Green
function. Therefore, the whole problem requires self--consistency.

In this paper we confine ourselves to zero temperature and make an
additional assumption on solving the impurity model using the Hartree--Fock
approximation. In this approximation the self--energy reduces to 
\begin{equation}
\Sigma_{\gamma_1\gamma_2}=\sum_{\gamma_3\gamma_4} (U_{\gamma _{1}\gamma
_{2}\gamma _{3}\gamma _{4}}-U_{\gamma _{1}\gamma _{2}\gamma _{4}\gamma
_{3}}) \bar{n}_{\gamma _{3}\gamma _{4}}  \label{HF}
\end{equation}
where $\bar n_{\gamma_1\gamma_2}$ is the average occupation matrix for the
correlated orbitals. The off-diagonal elements of the occupancy matrix are
not zero when spin-orbit coupling is included~\cite{sol}. The latter can be
implemented following the prescription of Andersen~\cite{OA75} or more
recent one by Pederson~\cite{pk}.

In the Hartree--Fock limit the self--energy is frequency independent and
real. The self--consistency condition of DMFT can be expressed in terms of
the average occupation matrix: Having started from some $\bar n%
_{\gamma_1\gamma_2}$ we find $\Sigma_{\gamma_1\gamma_2}$ according to~(\ref
{HF}). Fortunately, the computation of the on--site Green function~(\ref
{dmft}) needs {\em not} to be performed. Since the self--energy is real, the
new occupancies can be calculated from the eigenvectors of the one--electron
Hamiltonians with $\Sigma_{\gamma_1\gamma_2}-v_{dc}$ added to its $dd$
block. The latter can be viewed as an orbital--dependent potential which has
been introduced by the LDA+U method \cite{anisimov}.

The LDA$+$U method has been very successful compared with experiments at
zero temperature in ordered compounds. By establishing its equivalence to
the static limit of the DMFT we see clearly that dynamical mean--field
theory is a way of improving upon it, which is crucial for finite
temperature properties.

In this work we study the effect of the Slater parameters $F_{0}$, $F_{2}$
and $F_{4}$ on the magnetic anisotropy energy. Slater integrals can be
linked to intra--atomic repulsion $U$ and exchange $J$ obtained from LSDA
supercell procedures via $U=F^{0}$ and $J=(F^{2}+F^{4})/14$. The ratio $%
F^{2}/F^{4}$ is to a good accuracy a constant $\sim 0.625$ for $d$
electrons~ \cite{afl}. The MAE is calculated by taking the difference of two
total energies with different directions of magnetization (MAE=$E(111)-E(001)
$). The total energies are obtained via fully self consistent solutions.
Since the total energy calculation requires high precision, full potential
LMTO method~\cite{Sav} has been employed. For the $\vec{k}$ space
integration, we follow the analysis given by Trygg and co--workers~\cite
{tjew} and use the special point method~\cite{froyen} with a Gaussian
broadening~\cite{mp} of $15\ mRy$. The validity and convergence of this
procedure has been tested in their work ~\cite{tjew}. For convergence of the
total energies within desired accuracy, about $15000\ k$-points are needed.
We used $28000\ k$-points to reduce possible numerical noise, where the
convergency is tested up to $84000k$-points. Our calculations include
non-spherical terms of the charge density and potential both within the
atomic spheres and in the interstitial region~\cite{Sav}. All low-lying
semi-core states are treated together with the valence states in a common
Hamiltonian matrix in order to avoid unnecessary uncertainties. These
calculations are spin polarized and assume the existence of long-range
magnetic order. Spin-orbit coupling is implemented according to the
suggestions by Andersen~\cite{OA75}. We also treat magnetization as a
general vector field, which realizes non-collinear intra-atomic nature of
this quantity. Such general magnetization scheme has been recently discussed
\cite{singh}.

To incorporate the effects of intraatomic correlations on the
magnetocrystalline anisotropy energy, we have to take into account the
intra--atomic repulsion $U$ and the intraatomic exchange $J$. It is
important to perform the calculations for fixed values of magnetic moments
which themselves show a dependency on $U$ and $J$. Since the pure LSDA
result ($U$=0, and $J$=0) reproduces the experimental values for magnetic
moments in both Fe and Ni fairly well, we have scanned the $U-J$ parameter
space and have obtained the path of $U$ and $J$ values which hold the
theoretical moment constant, following the approach of Ref. \cite{Kudrnovsky}%
.

We now discuss our calculated MAE. We first test our method in case of LDA ($%
U=J=0$). To compare with previous calculations, we turn off the
non-collinearity of magnetization which makes it collinear with the
quantization axis. The calculation gives correct orders of magnitude for
both fcc Ni and bcc Fe but with the wrong easy axis for Ni, which is the
same result as the previous result~\cite{tjew}. Turning on the
non-collinearity results in a a larger value of the absolute value of the
MAE ($2.9\ \mu eV$) for Ni but the easy axis predicted to be (001) which is
still wrong. The magnitude of the experimental MAE of Ni is $2.8\ \mu eV$
aligned along $(111)$ direction~\cite{landolt}.

We now describe the effect of correlations, which is crucial in predicting
the correct axis of Ni (see Fig.\ \ref{mae}). We walked along the path of
parameters $U$ and $J$ which hold the magnetic moment to 0.6 $\mu _{B}$. The
MAE first increases to $60\ \mu eV$ ($U=0.5\ eV$, $J=0.3\ eV$) and then
decreases. While decreasing it makes a rather flat area from $U=1.4\ eV$, $%
J=0.9\ eV$ to $U=1.7\ eV$, $J=1.1\ eV$ where MAE is positive and around $10\
\mu eV$. After the flat area, the MAE changes from the wrong easy axis to
the correct easy axis. The correct magnetic anisotropy is predicted at $U
=1.9\ eV$ and $J=1.2\ eV$. The change from the wrong easy axis to the
correct easy axis occurs over the range of $\delta U\sim 0.2eV$, which is of
the order of spin-orbit coupling constant ($\sim 0.1eV$).

For Fe, the MAE is calculated along the path of $U$ and $J$ values which
fixes the magnetic moment to $2.2\ \mu _{B}$. At $U=0\ eV$ and $J=0\ eV$,
the MAE is $0.5\ \mu eV$. It increases as we move along the contour in the
direction of increasing $U$ and $J$. The correct MAE with the correct
direction of magnetic moment is predicted at $U=1.2\ eV$ and $J=0.8\ eV$.

Notice that the trends in the values of $U$ and $J$ that are necessary to
reproduce the correct magnetic anisotropy energy within LDA + U are similar
to the values used to describe photoemission spectra of these materials~\cite
{kl99} within DMFT. The values of the parameters $U$ and $J$, are basis--set
dependent, and method dependent, but the values of $U$ used in our LDA+ U
calculation are within $1eV$ of those used in \cite{kl99}. Since DMFT
contains the graphs which screen the on--site interaction which are ommitted
in the LDA+ U functional, a larger value of $U$ is needed to produce the
correct moment in DMFT.

We find direct correlation between the dependency of the MAE as a function
of $U$-$J$ and the difference of magnetic moments ($\Delta m=-(m(111)-m(001)$%
) behaving similarly (see Fig.\ \ref{mae}). For Ni the difference increases
till $U=0.4\ eV$ and $J=0.2\ eV$, then decreases. While decreasing it makes
a flat area from $U=0.9\ eV$ and $J=0.6$ to $U=1.7\ eV$ and $J=1.1\ eV$.
After the flat area, the difference decreases rapidly. For Fe, the
difference of magnetic moments slightly fluctuates till $U=0.7\ eV$ and $%
J=0.5\ eV$ and then decreases till $U=1.0\ eV$ and $J=0.7\ eV$ .

This concurrent change of MAE and the difference of magnetic moments
suggests why some previous attempts based on force theorem~\cite{dks} failed
in predicting the correct easy axes. Force theorem replaces the difference
of the total energies by the difference of one--electron energies. In this
approach, the contribution from the slight difference in magnetic moments
does not appear and, therefore, is not counted in properly. Unfortunately,
we could not find any experimental data of magnetic moments with different
orientations to the desired precision ($10^{-4}\mu_B$) to compare with.

We now present implications of our results on the calculated electronic
structure for the case of Ni. One important feature which emerges from the
calculation is the absence of the $X_2$ pocket (see Fig.\ \ref{fermi}). This
has been predicted by LDA but has not been found experimentally~\cite{wc}.
The band corresponding to the pocket is pushed down just below the Fermi
level. This is expected since correlation effects are more important for
slower electrons and the velocity near the pocket is rather small. It turns
out that the whole band is submerged under the Fermi level. We also find
that the removal of the $X_2$ point is near the point $U=1.9\ eV$ and $%
J=1.2\ eV$. For comparison, the corresponding band is just above the Fermi
level at $U=1.9\ eV$ and $J=1.1\ eV$ forming a tiny pocket. This strengthens
the connection between MAE and the absence of $X_2$ pocket.

There has been some suspicions that the incorrect position of the $X_2$ band
within LDA was responsible for the incorrect prediction of the easy axis
within this theory. Daalderop and coworkers~\cite{dks} removed the $X_2$
pocket by increasing the number of valence electrons and found the correct
easy axis. We therefore conclude that the absence of the pocket is one of
the central elements in determining the magnetic anisotropy, and there is no
need for any ad-hoc adjustment within a theory which takes into account the
correlations.

We now describe the effects originated from (near) degenerate states close
to the Fermi surface. These have been of primary interest in past analytic
studies~\cite{Kondorskii,mori:74}. We will call such states {\em degenerate
Fermi surface crossing} (DFSC) states. The contribution to MAE by non-DFSC
states comes from the fourth order perturbation. Hence it is of the order of 
$\lambda^4$. The energy splitting between DFSC states due to spin-orbit
coupling is of the order of $\lambda$ because the contribution comes from
the first order perturbation. Using linear approximation of the dispersion
relation $\epsilon(\vec{k}\lambda)$, the relevant volume in $k$-space was
found of the order $\lambda^3$. Thus, these DFSC states make contribution of
the order of $\lambda^4$. Moreover, there may be accidentally DFSC states
appearing along a line on the Fermi surface, rather than at a point. We have
found this case in our LDA calculation for Ni. Therefore the contribution of
DFSC states is as important as the bulk non-DFSC states though the
degeneracies occur only in small portion of the Brillouin zone.

%%%%%%%%%%%%%%%%%%%%%%%%%%%%%%%%%%%

The importance of the DFSC states leads us to comparative analysis of the
LDA and LDA+U band structures near the Fermi level. In LDA, five bands are
crossing the Fermi level at nearly the same points along the $\Gamma X$
direction. Two of the five bands are degenerate for the residual symmetry
and the other three bands accidentally cross the Fermi surface at nearly the
same points. There are two $sp$ bands with spin up and spin down,
respectively. The other three bands are dominated by $d$ orbitals. In LDA$+$%
U, one of the $d$ bands is pushed down below the Fermi surface. The other
four bands are divided into two degenerate pieces at the Fermi level (see
Fig. \ref{fermi}): Two symmetry related degenerate $d_\downarrow$ bands and
two near degenerate $sp_\uparrow$ and $sp_\downarrow$ bands. In LDA, we
found that two bands are accidentally near degenerate along the line on the
Fermi surface within the plane $\Gamma X L$. One band is dominated by $%
d_\downarrow$ orbitals. The other is dominated by $d_\downarrow$ orbitals
near $X$ and by $s_\downarrow$ orbitals off $X$. In LDA+U, these accidental
DFSC states disappear(see Fig. \ref{fermi}).

As we have seen, the on-site repulsion $U$ reduces the number of DFSC states
along $\Gamma X$ direction. Based on the tight--binding model, the
importance of DFSC states has been shown. We see that strong correlations
reduce number of DFSC states in $\Gamma X$ direction and remove the near
degenerate states on $\Gamma X L$ plane. We conclude that the change of DFSC
states is another important element that determines the easy axis of Ni.

%%%%%%%%%%%%%%%%%%%%%%%%%%%%%%%%%%%%%%%%%%%%%%%%%%%%%%%%%%%%%%%%%%%%%

%\section{Conclusion}

To conclude, we have demonstrated that it is possible to perform highly
precise calculation of the total energy in order to obtain both the correct
easy axes and the magnitudes of MAE for Fe and Ni. This has been
accomplished by including the strong correlation effects via taking
intra--atomic repulsion and exchange into account and incorporating the
non--collinear magnetization. In both Fe and Ni, both $U$ and $J$ take
physically acceptable values consistent with the values known from atomic
physics. The calculations performed are state of the art in what can
currently be achieved for realistic treatments of correlated solids. Further
studies should be devoted to improving the quality of the solution of the
impurity model within DMFT and extending the calculation to finite
temperatures.

This research was supported by the ONR grant N0014-99-1-0653. GK would like
to thank K. Hathaway for discussing the origin of magnetic anisotropy and G.
Lonzarich for discussing dHvA data. We thank R. Chitra for stimulating
discussion. We thank V. Oudovenko for setting up the computer cluster used
to perform these calculations. We have also used the supercomputer at the
Center for Advanced Information Processing, Rutgers. IY thanks K. H. Ahn for
discussions.

%\bibliography{slda}

\begin{references}
\bibitem{vanVleck:37}  J.~H. van Vleck, Phys.\ Rev.\ {\bf 52}, 1178 (1937).

\bibitem{brooks}  H. Brooks, Phys.\ Rev.\ {\bf 58}, 909 (1940).

\bibitem{fletcher}  G.~C. Fletcher, Proc.\ R.\ Soc.\ London {\bf 67A}, 505
(1954).

\bibitem{sloncewskij}  J.~C. Sloncewskij, J.\ Phys.\ Soc.\ Jpn.\ {\bf 17},
Suppl.\ B (1962).

\bibitem{asdente}  M. Asdente and M. Delitala, Phys.\ Rev.\ {\bf 163}, 497
(1967).

\bibitem{hk64}  P. Hohenberg and W. Kohn, Phys.\ Rev.\ {\bf 136}, 864 (1964).

\bibitem{ks65}  W. Kohn and L.~J. Sham, Phys.\ Rev.\ {\bf 140}, 1133 (1965).

\bibitem{efn}  H. Eckardt, L. Fritsche, and J. Noffke, J.\ Phys.\ F {\bf 17}%
, 943 (1987).

\bibitem{halilov}  S.~V. Halilov {\it et~al.}, Phys.\ Rev.\ B {\bf 57}, 9557
(1998).

\bibitem{jansen}  H.~J.~F. Jansen, J.\ Appl.\ Phys.\ {\bf 67}, 4555 (1990).

\bibitem{tjew}  J. Trygg, B. Johansson, O. Eriksson, and J.~M. Willis,
Phys.\ Rev.\ Lett.\ {\bf \ 75}, 2871 (1995).

\bibitem{dks}  G.~H.~O. Daalderop, P.~J. Kelly, and M.~F.~H. Schuurmans,
Phys.\ Rev.\ B {\bf \ 41}, 11 919 (1990).

\bibitem{freeman}  D. Wang, R. Wu, and A.~J. Freeman, Phys.\ Rev.\ Lett.\ 
{\bf 70}, 867 (1993).

\bibitem{beiden}  S.~V. Beiden, W.~M. Temmerman, Z. Szotek, G.~A. Gehring,
G.~M. Stocks, Y. Wang, D.~M.~C Nicholson, W.~A. Shelton, and H. Ebert,
Phys.\ Rev.\ B {\bf 57}, 14 247, (1998).

\bibitem{anisimov}  {For a review, see, e.g., {\em Strong Correlations in
electronic structure calculations}}, edited by V.~I. Anisimov (Gordon and
Breach Science Publishers, Amsterdam, 2000).

\bibitem{singh}  L. Nordstrom and D. Singh, Phys.\ Rev.\ Lett.\ {\bf 76},
4420 (1996).

\bibitem{gabi:rmp}  A. Georges, G. Kotliar, W. Krauth, and M. Rozenberg,
Rev.\ Mod.\ Phys.\ {\bf \ 68}, 13 (1996).

\bibitem{laz}  A.~I. Liechtenstein, V.~I. Anisimov, and J. Zaanen, Phys.\
Rev.\ B {\bf 12}, 3060 (1975).

\bibitem{OA75}  O.~K. Andersen, Phys.\ Rev.\ B {\bf 12}, 3060 (1975).

\bibitem{sol}  I.~Y. Solovyev, A.~I. Liechtenstein, and K. Terakura, Phys.\
Rev.\ Lett.\ {\bf \ 80}, 5758 (1999).

\bibitem{pk}  M.~R. Pederson and S.~N. Khanna, Phys.\ Rev.\ B {\bf 60}, 9566
(1999).

%\bibitem{laz}
%A.~I. Liechtenstein, V.~I. Anisimov, and J. Zaanen, Phys.\ Rev.\ B {\bf 12},
%  3060  (1975).

\bibitem{afl}  V.~I. Anisimov, F. Aryastiawan, and A.~I. Lichtenstein, J.\
Phys.: Condensed Matter {\bf 9}, 767 (1997).

\bibitem{Sav}  S.~Y. Savrasov, Phys.\ Rev.\ B {\bf 54}, 16470 (1996).

\bibitem{froyen}  S. Froyen, Phys.\ Rev.\ B {\bf 39}, 3168 (1989). A.~P.
Cracknell, J.\ Phys.\ C {\bf 2}, 1425 (1969).

\bibitem{mp}  R.~J. Needs, R.~M. Martin, and O.~H. Nielsen, Phys.\ Rev.\ B 
{\bf 33}, 3778 (1986). K.-M. Ho, C.~L. Fu, and B.~N. Harmon, Phys.\ Rev.\
Lett.\ {\bf 49}, 673 (1982).

\bibitem{Kudrnovsky}  M. Pajda, J. Kudrnovsky, I. Turek, V. Drchal, P.
Bruno, cond-mat/0007441;

\bibitem{landolt}  M.~B. Stearns, {in {\em Magnetic Properties of $3d$, $4d$%
, and $5d$ Elements, Alloys and Compounds}}, Vol.\@ III of {\em %
Landolt-B\"ornstein, New Series}, edited by K.-H. Hellewege and O. Madelung,
(Springer-Verlag, Berlin, 1987).

\bibitem{kl99}  M. Katsenelson and A. Lichtenstein, J.\ Phys.\ Cond.\ Matt.\ 
{\bf 11}, 1037 (1999). M. Katsenelson and A. Lichtenstein, Phys.\ Rev.\ B.\ 
{\bf 61}, 8906 (2000).

\bibitem{wc}  C.~S. Wang and J. Callaway, Phys.\ Rev.\ B {\bf 9}, 4897
(1973). F. Weling and J. Callaway, Phys.\ Rev.\ B {\bf 26}, 710 (1982).

\bibitem{Kondorskii}  E.~I. Kondorskii and E. Straube, Sov.\ Phys.\--JETP 
{\bf 36}, 188 (1973).

\bibitem{mori:74}  N. Mori, Y. Fukuda, and T. Ukai, J.\ Phys.\ Soc.\ Jpn.\ 
{\bf 37}, 1263 (1974).
\end{references}
%\bibliographystyle{prsty}

%%%%%%%%%%%%%%%%%%%%%%%%%%%%%%%%%%%%%%%%%%%%%%%%%%%%%%%%%%%%%%%%%%%%%%%%%%%%%
%%%%%%
%%%%%%           Figure CAPTIONS
%%%%%%
%%%%%%%%%%%%%%%%%%%%%%%%%%%%%%%%%%%%%%%%%%%%%%%%%%%%%%%%%%%%%%%%%%%%%%%%%%%%%
%----------------------------------------------------------------------------

\begin{center}
\begin{figure}[tbp]
\caption{The magnetocrystalline anisotropy energy $\mbox{MAE}=E(111)-E(001)$
(circle) and the difference of magnetic moment $\Delta m=m(001)-m(111)$
(square) for Ni (top) and for Fe (bottom) as functions of $U$. The
experimental MAEs are marked by arrows for Fe ($1.4\ \protect\mu eV$) and Ni
($-2.8\ \protect\mu eV$). The values of exchange parameter $J$ for every
value of $U$ are chosen to hold the magnetic moment of 0.6 $\protect\mu _B$
in Ni and 2.2 $\protect\mu _B$ in Fe}
\label{mae}
\end{figure}
\end{center}

%----------------------------------------------------------------------------

%----------------------------------------------------------------------------

\begin{center}
\begin{figure}[tbp]
\caption{Calculated Fermi Surface of Ni with the correlation effects taken
into account. The solid and dotted lines correspond to majority and minority
dominant spin carriers. Dominant orbital characters are expressed. Both
experimentally confirmed $X_5$ pocket and $L$ neck can be seen. The $X_2$
pocket is missing, which is in agreement with experiments.}
\label{fermi}
\end{figure}
\end{center}

%----------------------------------------------------------------------------

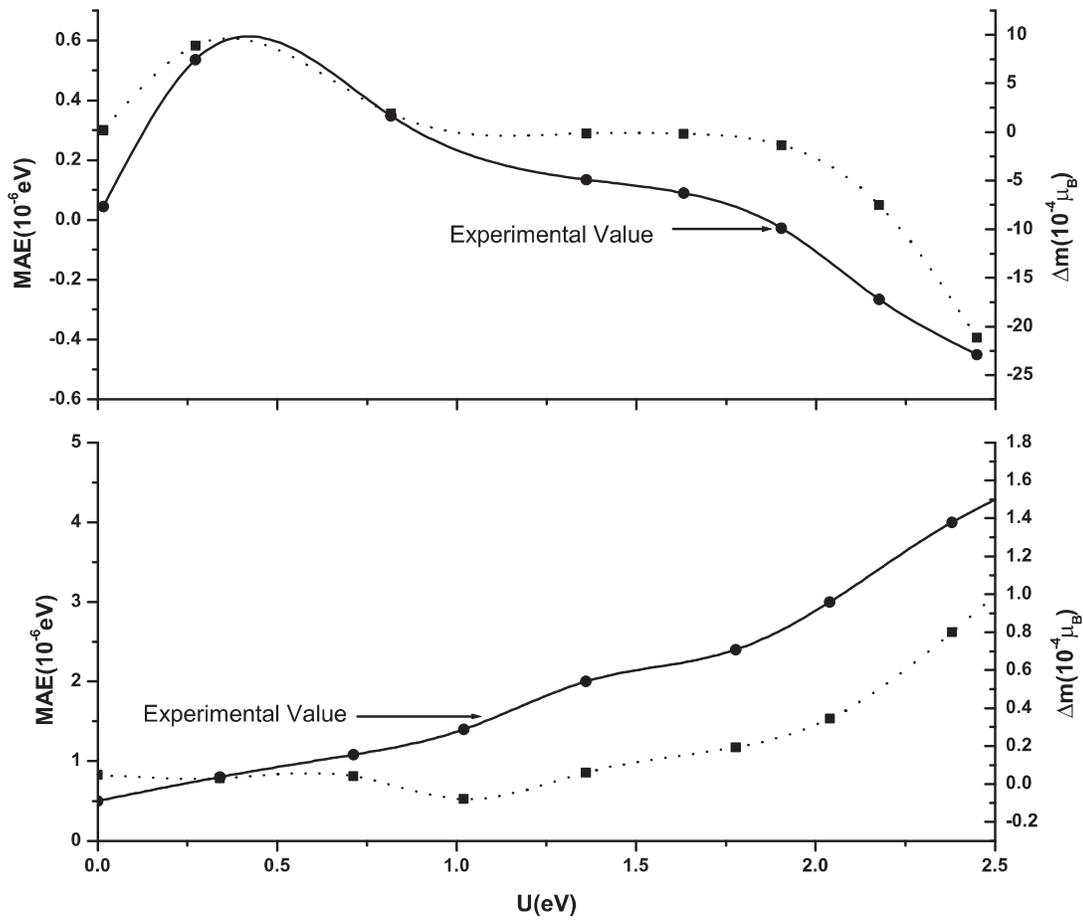
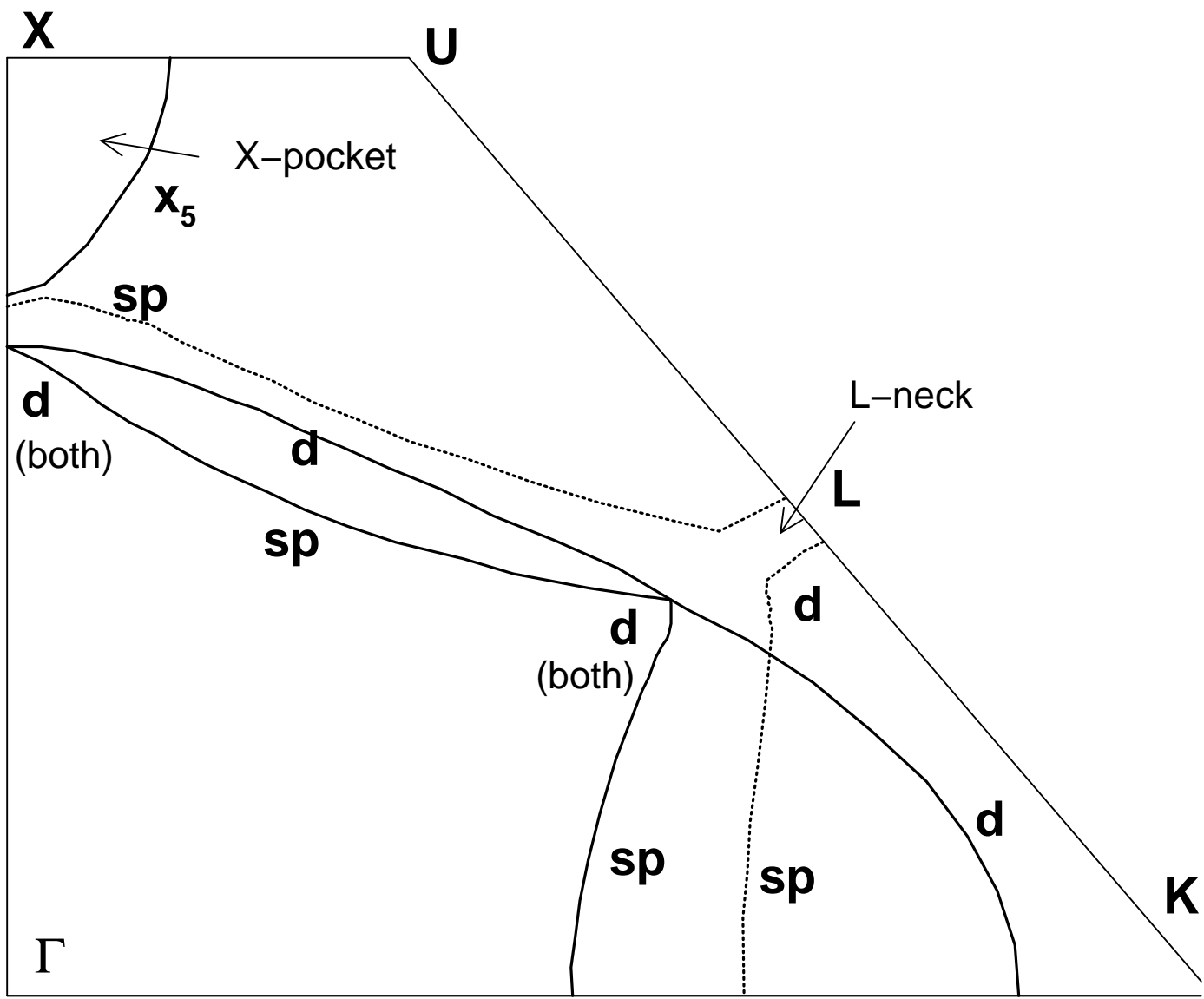
\end{document}